\documentclass[preprint,showpacs,preprintnumbers,amsmath,amssymb,floats]{revtex4}
\usepackage{blindtext}
\usepackage{graphicx}
\usepackage[english]{babel}
\usepackage{amsmath}
\usepackage{bm}
\usepackage{amssymb}
\usepackage{color}

\usepackage[mathletters]{ucs}
\usepackage[utf8x]{inputenc}

\newcommand{\be}{\begin{eqnarray}}
\newcommand{\ee}{\end{eqnarray}}
\usepackage[toc,page]{appendix}

\newcommand{\hide}[1]{{}}

\begin{document}

\title{ ${\bf \frac{h}{e}}$~flux quantization in metals due to Berry phase coherence }
\author{Chandra M. Varma}
\affiliation{150 Thompson Street, New York, NY 10012}
\affiliation{\footnote{Emeritus Distinguished Professor}Department of Physics, University of California, Riverside, CA. 92521, USA}
\date{\today}

\begin{abstract}
Berry curvature does not show itself in the relative phase correlation of wave-functions at different spatial points  in a metal unless the fermions have closed trajectories in momentum space, for example those around isolated impurities. But these, just as the Bloch phase correlations, disappear at lengths larger than the diffusion length. If a  quasi-two dimensional metal with Berry curvature has a set of domains, their boundaries necessarily carry chiral currents precluding back-ward scattering.  The Berry induced phase coherence then persists over length scales of order the scale at which the chiral one-dimensional states scatter into the bulk states, which can be macroscopic. The conditions for their occurrence and  the lengths and the orientation of such states are derived. These calculations are used to understand the remarkable aspects of a recent experiment in an anisotropic metal, reported to have loop-current order, with mean-free path of about 0.01 $\mu$m which exhibits flux quantization in some  transport properties over lengths of several $\mu$m. 

\end{abstract}

\maketitle
{\it (i) Introduction}: Phase coherence in the usual metallic state, except due to an external magnetic field,  can occur  only on length scales much smaller than the mean-free path. This work is motivated by an experiment \cite{chunyu2025longrangeelectroncoherencekagome} which requires macroscopic phase coherence in the static one-particle fermion correlations. The experiment observes, below  a temperature $T_{\ell} \approx 35$ K, a  periodic contribution to the c-axis magneto-resistance 
 with flux quantum $\phi_0 = h/e$, for fields applied $B$ parallel to the  a-b layers of the  kagome 
compound CsV$_3$Sb$_5$ (CVS). 
The resistive mean free path $\ell_{mf}$ near $T_{\ell}$ is about $10^{-2} \mu m$, much smaller than the size of the sample.
 The geometry, see Fig. (1), is akin to that of the Fraunhofer pattern with $h/2e$ flux quantization in Josephson junctions of weakly coupled superconductors separated by $d$, except that in the present problem, the resistance  is measured over a $\mu$m distance in the c-direction, across about $10^3$ layers.
 
 CVS  has shown evidence \cite{Hasan2022, xing2024opticalmanipulationchargedensity, Guo_2024, wei2024threedimensionalhiddenphaseprobed, Wilson2023, Li_ge_2024, Mielke2022, Xiang_2021, LeTian2024, ge2025nonreciprocalsuperconductingcriticalcurrents, varmanrcc2025}
  below a temperature (which appears to vary between samples) of a Berry curvature \cite{Berry1984} due to loop-current order \cite{cmv1997, SimonV2002, Weber-Giam-V}. Such orders are promoted by non-local interactions, primarily to reduce frustrations of multiple kinetic energy paths in models with multiple nearly degenerate orbitals in a unit-cell - see \cite{Shekhter2009} for a pedagogical paper. Various patterns of loop-currents  have been proposed for CVS \cite{feng2021chiral, lin2021loop, park2021theory, tazai2023charge, denner2021analysis, shumiya2021intrinsic, Nakazawa_2025},  but the experiments cannot yet be said to specify the symmetries beside that time-reversal and chirality are broken. 
 There are also experiments in some samples which do not see evidence for time-reversal breaking \cite{saykin2025highresolutionpolarkerr, Zeljkovic2025}. 
 
 \begin{figure}
 \begin{center}
 \includegraphics[width= 0.6\columnwidth]{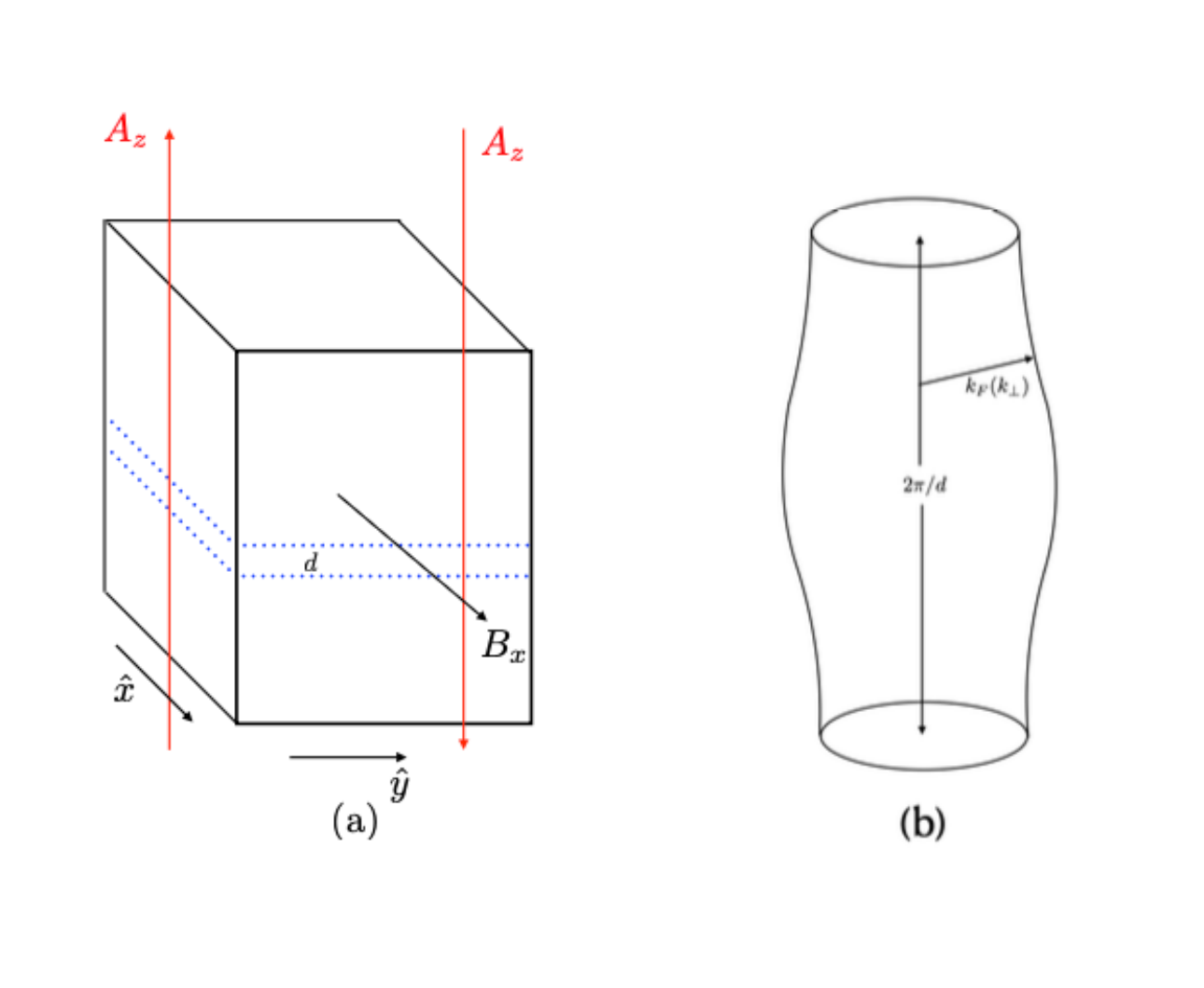}
 \end{center}
\caption{(a): Figure sketching the geometry of the magnetic field in the x-direction, and the choice of vector potential in Moll's experiment on stacks of two-dimensional layers. The stacks of parallel planes in the vertical direction are displaced by $d$. $A {\hat z}$ goes from $- B_x L_y/2$ at one end of the sample in the $y$-direction to $B_x L_y/2$ at the other end; $L_y$ is the width of the sample.
(b): Sketch of the simplest 3 dimensional Fermi-surface without flux.}
 \label{Fig:Geometry}
\end{figure}

{\it (ii) The experiments}: Beside the original source \cite{chunyu2025longrangeelectroncoherencekagome}, a concise summary of the experiments is given in \cite{Schmalianjccp2025}. Figure (\ref{Fig:Geometry}) shows Moll's stack of kagome planes with a magnetic field ${\bf B}$ parallel to the planes in the $\hat{\bf x}$ direction, which can be represented by a vector potential ${\bf A}$ and has a flux $\phi_x$ in the area $d L_y$,
 \be
\label{AB}
{\bf A}_z  ~\hat{\bf z} =  - B_x y~  \hat{\bf x},~~\phi_x \equiv B_x d L_y.
\ee
   The principal features of the experiments  are:\\
(1). In a  rectangular prism of cross-section $(L_x,L_y)$ with both about 2 $\mu$m, and height about 1$\mu$m, a magnetic field is applied parallel to the planes. c-axis resistivity in the field is measured. Beside the usual contribution $\propto B^2$, an oscillatory component  a period of the flux quantum $\phi_0 \equiv h/e$ is found for flux $\phi_x = B_x L_y d$ or $\phi_y = B_y L_x d$, with an abrupt switch between them as the angle of ${\bf B}$ in the plane is varied.\\
(2). The amplitude of the oscillation approaches zero  rapidly for fields perpendicular to the planes as small as about 100 Gauss.  The observed magneto-chirality in the same compound \cite{Guo_2024}, which also requires fields parallel to the planes also disappears with a similar perpendicular field. \\
(3). Only the component of the field $B_{x}$ is effective for $B_{x} > B_{y}$, and vice-versa. The period of the oscillations depends not on  $|B|$ but on $B_x$ or $B_y$. A fairly sharp switch occurs at $B_x \approx B_y$ for a rectangular sample. This is irrespective of the ratio $L_x/L_y$. This is the most astonishing feature of the results. \\

Below, each of these observations is understood on the basis of a model. In (iii) the lack of coherence of wave-functions beyond a diffusive length is re-capitulated; in (iv) a model of one-dimensional chiral domain walls  along which Berry phase coherence is retained is introduced; in (v) how this coherence is lost in magnetic fields perpendicular to the planes and at what value is derived; in (vi), given the coherence in the domain walls, the Hamiltonian for coherent propagation of fermions between adjacent layers, and the effect of a magnetic field parallel to the planes is introduced; in (vii) the energetics of the formation of domain walls is considered so that they are plausible; in (viii), the c-axis quantized magneto-resistance is derived and features such as (3) above are shown to follow; predictions for future experiments are suggested in (ix).

{\it (iii) Coherence of wave-functions in the bulk ?} : Let us consider the Fourier transform of the static one-particle correlation function
\be
G_{\bf k}({\bf r_1,r_2}) = \langle \psi^+({\bf r_1}) \psi({\bf r_2})e^{- i{{\bf k. (r_1-r_2)}}} \rangle,
\ee
where $\psi^+({\bf r}),  \psi({\bf r})$ are fermion creation and annihilation operators (spin is not relevant to the calculations). In the pure limit, the Berry phase can be shown to be absent in the phase factor of $G_{\bf k}({\bf r_1,r_2})$ for any ${\bf k}$ by requiring gauge invariance. The Berry phase makes its appearance only when there is interference in ${\bf k}$ space, due to scattering of  Bloch waves by impurities or a magnetic field perpendicular to the planes. This problem has been worked out for impurity scattering by Dutreix et al. \cite{Dutreix2021}. The answer for a single impurity and a magnetic field and for asymptotically large $r = |{\bf r_1-r_2}|$ at ${\bf k = k_F}$,
\be
\label{Dutr}
G_{k_F}({\bf r_1,r_2}) \propto \frac{1}{{\sqrt r}} e^{i \big(2 k_F r + \delta_F(E_F) + \phi_B(\theta) + \phi^{loop}_{AB}({\bf r})\big)}.
\ee
$\delta_F(E_F)$ is the Friedel phase due to the impurities,  $\phi_{AB}^{loop}$  is the Aharanov-Bohm (AB) phase, which is the integral of the vector potential over the closed loop formed by the path between ${\bf r}_1$ and ${\bf r}_2$ enclosing the impurity. $\phi_B(\theta)$, the phase we are most interested in, encodes the Berry phase.
 This naturally depends on the model. For a two band model with a spinor for $u_{\bf k}$ with a relative phase factor $e^{i \theta_{\bf k}}$ between the two components, the back-scattering overlap of the spinors (the most important scattering to form a closed loop) due to impurity scattering  is,
\be
\langle u_{-\bf k} | u_{\bf k}\rangle \propto e^{i\phi_B(\theta)}; ~~ \phi_B(\theta) = (\theta_{\bf k} - \theta_{\bf -k}).
\ee
For a density of impurities, one must sum over closed trajectories $\gamma$ between ${\bf r_1}$ and ${\bf r_2}$. In the adiabatic approximation,
\be
\label{Gav}
G({\bf r_1, r_2}, E_F) \approx \sum_{\gamma -{\bf r_1, r_2}} {\mathcal{\bf a}}_{\gamma}(E_F) e^{i \phi_{\gamma}(E_F)} e^{i\frac{({\bm{r}_1} \times  {\bm {r}_2})\cdot \hat{z}}{\ell_B^2}}.
 \ee
The last factor is the AB phase due to  the trajectory, which enforces the magnetic-periodicity. $\ell_B = \sqrt{\frac{\hbar}{2 eB_z}}$.
First consider the averaging without $B_z$. Then $\phi_{\gamma}$ includes only the Friedel contribution and  the Berry phase $\phi_B$ at the chemical potential in the trajectory $\gamma$.
For weak scattering with Gaussian random distribution, the average over the trajectories has been performed before in other contexts \cite{Hikami1980} and the presence of the Berry phase term does not cause additional technical difficulty. The Gaussian random average averages over products of two Green's functions over trajectories $\gamma$ and $\gamma', \langle G^R({\bf r_1, r_2})G^A({\bf r_1, r_2})\rangle$. This leads to summation over Diffusion propagators as well as Cooperons, see for example \cite{Hikami1980, Altshuler1981, NagaosaRMP2010}.  I only note that the phase factors (in the absence of a magnetic field) disappear. $G({\bf r_1, r_2, E})$ is a diffusive propagator with an extra dephasing term.  The final result  is that there can be no phase coherence due to Berry curvature if the mean-free path $\ell_{mf}$ is smaller than a Berry de-phasing length $L_{\Omega}$, which is set by the Bloch-space spinor geometry of the Fermi-surface. The Berry space de-phasing length $L_{\Omega}$, (see for example \cite{BerryDephasing}) is given by
\be
L_{\Omega} =  \sqrt{D ~ \tau_{\Omega}},~~ \frac{\tau_{\Omega}}{\tau}  = \ln \big \langle |\langle u_{\bf k}| u_{\bf k}' \rangle |^2\big\rangle_{{\bf k,k}' \subset {\bf k}_F}.
 \ee
 $\tau$ is the impurity scattering rate and $ D = v_F^2 \tau$. 
 So for distance much larger than  $L_{\bf \Omega}$, there is no phase coherence of the form that is necessary for the experiments of Moll et al. For a two band Haldane like model \cite{Haldane1988}, it depends on filling - but  $L_{\bf \Omega} \lesssim {\ell}$, generally. This kind of result has already been derived for the $\pi$ Berry phase of the Dirac cone in graphene, \cite{McCann_2013} and other contexts \cite{BerryDephasing}. 
   
{\it (iv) Coherence along one-dimensional chiral channels} :  All is not lost. One must consider the possibility of chiral domain walls embedded in the metal in loop-current models. They necessarily carry currents \cite{CMV2019} and separate domains with different Chern numbers (quantized or un-quantized). They must either form closed loops to conserve current or a network of chiral domain walls as proposed \cite{Chalker_1988} for the quantum-Hall effect plateau transition. The conditions for the occurrence of the requisite channels for our problem are derived in (vii) below.
 
   To begin with, only parallel chains without intersection are considered. The domain walls are one dimensional chiral channels, so elastic  back-scattering is not possible. A stable phase factor in $G({\bf r_1,r_2})$ along the channels may therefore be preserved, but limited by the fact that the channels are buried in the bulk with gapless excitations, so that there is scattering to diffusive fermions.  At finite temperature, there is also inelastic de-phasing. All this may be parametrized by a loss length $\ell_{1d}$.  
 The two point correlation on points on the chiral domain walls is
 \be
 \label{L1d}
 G_{net}({\bf r_1, r_2}) \approx \sum_{\gamma, {\bf r_1,\to r_2}} A_{\gamma} e^{i \big(k_{\gamma} L_{\gamma}+ \phi_B(\gamma) \big)} e^{- \frac{L_{\gamma}}{\ell_{1d}}}.
 \ee
$\gamma$ now sums over trajectories in the bulk with lengths $L_{\gamma}$ which end at ${\bf r_1}$ and ${\bf r_2}$ .  The calculation of the loss length $\ell_{1d}$ defined in Eq. (\ref{L1d}) depends on many details. Multiple physical reasons contribute, for example impurity scattering into bulk states of the same energy, and tunneling into bulk states if the wall is quite thick . The simplest calculation is given here which is a lower limit to the scattering rate. Quite generally, $\ell_{1d}$ depends on the wave-function on the domain wall, which may be specified by transverse length $\xi_{\bot}$ which gives how bound the current carrying state is transverse to it. For a two band model with velocity of the mode along the domain wall $v \hat{x}$ and gap $2 m_0$, the decay length in the $\hat{y}$ direction from a Jackiw-Rebbi type calculations \cite{Xie2025} is $\xi_{\bot} = v/m_0$. For reasonable values $v = 10^5 cm/sec, m_0 = 50 K, \xi_{\bot} \approx 0.01 \mu m$. 
 A perturbative calculation mixing the 1-d mode with a fixed chirality $\chi$ given by the wave-function for a straight segment of length $L$,
 \be
 \psi_{1d} = \frac{e^{i k_x x}}{\sqrt{L}} \phi_{\bot}(y) \chi
 \ee
 where $\phi_{\bot}(y)$ decays exponentially with length $\xi_{\bot}$ has a matrix element $t_{\bot}({\bf k})$ with the bulk state with the same energy through a disorder potential of the order of $m$. Assume a Bloch wave for the 2-d states with a spinor $u_{\bf k}$ and normalization in $\sqrt{L \times L}$, gives a matrix element per unit-length 
 \be
 \label{tperp}
 t_{\bot}({\bf k}) \approx m \frac{1}{\sqrt {L \xi_{\bot}}} <u_{\bf k}|\chi>.
 \ee
 $<u_{\bf k}|\chi>$ projects the chirality of a given domain wall to the Bloch wave-functions. Eq. (\ref{tperp}) can be used after averaging to calculate  $ \ell_{1d} = v_x \tau_{1d}$, with $\tau^{-1}_{1d}$, the decay rate 
 \be
 \label{rate}
 \tau^{-1}_{1d} \approx 2 \pi m^2 \nu_{2D}(E_F) \sqrt{\frac{L}{\xi_{\bot}}}|<u_{\bf k}|\chi>|^2.
 \ee
 For the  values for $m$ and $v_y$which gave $\xi_{\bot} \approx 0.01 \mu m$, and  $\nu_{2D} \approx (1 eV)^{-1}, |<u_{\bf k}|\chi>|^2 \approx 1/10$, $L \approx 1 \mu m$, this gives $\ell_{1d} \approx 5~\mu m$. If one considers tunneling from the 1d-channel to the continuum, a much longer length is obtained, and so also if the matrix element is through fairly dense defects along the domain wall with coupling-energies much larger than $m$. Eq. (\ref{L1d}) may therefore be the upper limit to the requirement for coherence of the Berry phase over the requisite length of the domain wall, which  is expected to have a typical length of the sample size of a few $\mu m$. A lower estimate of the size of domain area  is obtained below by considering the observed destruction of the coherence by a magnetic field perpendicular to the planes.

{\it (v) Removal of coherence even in 1d channels by a magnetic field perpendicular to the planes}: 
  Let us now restore the AB phase in Eq (\ref{Gav}). The effect of disorder scattering in removing the AB phase correlations is quite benign compared to  removing the Berry phase correlations.  We now derive the scale at which $\phi_{AB}$ phase correlations are also lost due to disorder.
  It is evident from the correlation Eq. (\ref{Gav}) that $\phi_{AB}(\gamma)$ and $\phi_{B}$ are not independent and  that if the AB coherence is lost in the bulk, so is the Berry phase coherence in the 1d walls. From Eq. (\ref{Gav}), one must consider the statistically average over trajectories $\gamma$ with closed loops $\gamma$ over the bulk, with  ${\bf r}_1, {\bf r}_2$ on the domain walls.  
  A given trajectory has a fixed $\phi_{AB}^{\ell}$, but between ${\bf r}_1$ and ${\bf r}_2$, it gets de-phased in the sum over loops of random area. The typical area of the loop is of $O((L_{\gamma})^2)$. Unless domains  are very regular, the spread of the areas $\Delta A$ is of the same order. The AB phases between alternative trajectories de-cohere, when
 \be
 \label{Bc}
 2 \pi \frac{B_{z} \Delta A}{\phi_0} \gtrsim 1.
 \ee
 This is the condition that when disorder is such that the root mean-square deviation in an area enclosing $\phi_0$ is itself $\phi_0$, no effect of the magnetic periodicity is maintained. 
We can estimate from (\ref{Bc})  there is no AB coherence  and therefore no Berry phase coherence 
at fields of  of O(1 milli-Tesla) for length scales larger than O(1 $\mu$-m). This is in effect the length over which a disordered Hofstadter pattern of states removes phase coherence.

 {\it (vi) Coherent propagation between planes across the coherent chains}: For a magnetic field $B_x$ in the plane and with a gauge as in Eq. (\ref{AB}), the transfer integral $t_c$ from one plane to another for nearest neighbor atoms in the c-direction acquires a $y$-dependent Peierls phase for $-L_y/2 < y <L_y/2$
\be
t_{c}(y) \to t_{c}e^{i \big(\frac{2\pi \phi}{\phi_0} \frac{y}{L_y}\big)}. 
\ee
The fermi-surface may be assumed open in the transverse direction. For simplicity, we assume it to be simply connected circular in the planes, as shown in Fig.(\ref{Fig:Geometry}).  The three-dimensional occupied Fermi-surface for $\phi = 0$ may be described as in the figure by the two-dimensional Fermi-vector ${\bf k}_F(k_{c})$; $k_{c}$ being the transverse momentum. The dispersion in the c-direction is,
\be
\label{ep}
\epsilon_{\bot} (k_{c}) = - 2t_{c} \cos (k_{c} d), ~ -\pi/d < k_{c} < \pi/d.
\ee
The Peierls phase of the coherent channels has a  shift of phase in the wave-functions because of which  the momentum in the c-direction is in the gauge chosen,
\be
\label{shift}
k_{c}(y) d = k^0_{c} d + \frac{2\pi B_x d y}{\phi_0},
\ee
 Thus for any $y$ and $\phi$, the dispersion in the direction perpendicular to the planes  is given by that of an incommensurate lattice.
The velocity at the Fermi-surface in the c-direction is
  \be
 v_{Fc} (y) = 2t_{c} d \sin \big(k^0_{c} d + 2\pi \frac{B_x d y}{\phi_0}\big).
 \ee
  
 {\it (vii) Energetics of chain formation promoted by inter-chain coherent tunneling}: Let us first consider $B=0$. The coherent tunneling inter-layer Hamiltonian and the energy reduction due to it over macroscopic lengths occurs only if the wave-functions in each layer are coherent over such lengths, i.e. only between chains in successive layers. For a pair of such layers, the energetic cost of producing  a pair of such chains for chains of width $\xi_{\bot}$ in a length $L$  and width $\xi_{\bot}$ is given by~
$2  \kappa  (a_0/\xi_{\bot}) \xi_{\bot} L$.
  $\kappa$ is the stiffness energy of the loop-current order parameter in a unit-cell of length $a_0$. The energy gain between the two chains due to coherent tunneling depends on details such as the shape of the 3d Fermi-surface, but generically is of $- O(t_c \xi_{\bot} L)$. So for $t_c >> 2 \kappa$, the formation of coherent chains is promoted. Typically, in the pure limit $t_c$ from band-structure calculations is expected to be $\approx 0.1$ eV. We expect the stiffness (in the absence of any precise knowledge) to be similar to the energy reduction due to loop-current order, i.e. of $O(0.01 eV)$ per unit-cell. So given that $a/\xi_{\bot} \approx 10$, the formation seems plausible. In the absence of coherence (in the rest of the region of the layers, the characteristic energy gain due to tunneling is only $- O(t_c^2 \nu_{2d})$, which is at least an order of magnitude smaller than $-t_c$, because $\nu_{2d}$ is the two-dimensional density of states. 
 
Suppose there are only parallel chiral channels; they must be farther than $2\xi_{\bot}$  from each other. Assuming that they are repulsive, they are $O(4 \xi_{\bot})$ apart. This gives that they occupy a fractional area in the plane which is $O(\xi_{\bot}/{4 L})$. Note that since currents are not allowed perpendicular to the sample boundaries, chiral current-carrying channels can only flow at the boundaries of the sample only along the boundaries. Conservation requires that alternate parallel channels have opposite chirality; the change must happen at the transverse boundary on lengths of strips of $O(4 \xi_{\bot})$. This contribution need not concern us as it will give an energy proportional to $L_x$ or $L_y$, whereas the rest is proportional to the area.

If the chiral channels are not parallel, they must intersect. I now show that intersections are forbidden. The Hamiltonian-density for chiral channels in the x and in the y directions with densities $n_x, n_y$ per unit-length, with continuum wave-functions $\psi_x, \psi_y$, and with intersections at points $i$ 
\be
H_{x,y} = -i v \psi_x^+\partial \psi_x - i v \psi_y^+\partial \psi_y + \lambda \psi_x^+(i) \psi_y(i) + h.c.
\ee
$\lambda$ represents the finite interaction energy, and has the dimensions of energy $\times$ length. The intersection density of points $i$ is $n_x n_y$ per unit-area. Each intersection contributes a scattering rate $\approx |\lambda|^2/v$. So the total dephasing rate due to intersections is $\tau_{i.s.}^{-1} \approx n_xn_y |\lambda|^2/v$. The  length over which coherence of a channel is retained is $\ell_{i-s} = v \tau_{i-s}$. The energy per unit-length  with intersections due to coherent transfer from a plane to its neighbors including the energy cost considered above for making the 1d-channels per unit-length is then
\be
E_{i-s} \approx \kappa (n_x +n_y) - t_c \frac{v^2}{|\lambda|^2} \frac{1}{n_xn_y}.
\ee
It is easily seen that this is a winner take all situation; there is no stable minima except $n_x =0$ or $n_y=0.$ (Actually the minimum of $n_x, n_y$ is
of-course $1/L_x, 1/L_y$.)
 
{\it (viii) Magneto-resistivity in c-direction}:  With a magnetic field in the plane, the coherent transfer of fermions between the planes is periodically frustrated due to the Peierls phase shift (\ref{shift}). Let us calculate the eigenvalues (\ref{ep}) as a function of angle $\theta$ of ${\bf B}$ with respect to the $x$-axis by generalizing Eq. (\ref{shift}) and integrating over $x$ and $y$ over the sample. The energy at a fixed $k^0_c$ is calculated as 
 \be
 \label{En}
\overline{\epsilon_{\bot}}(k_c) = -2t_c \cos(k^0_c d) R(\phi_x, \phi_y), ~~
 R(\phi_x, \phi_y) &=& \Big(\frac{ \sin (\pi \frac{B_x d L_y}{\phi_0})}{\frac{\pi B_x d L_y}{\phi_0}}\Big)\Big( \frac{ \sin (\pi \frac{B_y d L_x}{\phi_0})}{\frac{\pi B_y d L_x}{\phi_0}}\Big).
  \ee
  $R(\phi_x, \phi_y)$ is a product of two functions, one for chiral channels in the $\pm x$ directions, the other for $\pm y$ directions. Using what has been shown above, only one of them is functional while the other gives $1$. Calculating the energy by integrating over  periods in $\phi_x/\phi_0$ or $\phi_y/\phi_0$ , the first term wins for $B_x/B_y >1$ and vice-versa. The energies are equal at $B_x =B_y$ irrespective of $L_x/L_y$. From this, the periodicity of the magneto-resistive flux quantization for $ \phi_{x,y}/\phi_0 =1$ and the switch, discussed in experimental facts (1) and (3) follow. 
   
 In considering the in-plane  resistivity in CVS, we can disregard the conductivity of the network of quasi-one dimensional chains  both because it occupies a small fraction of the sample, and because the current has to travel to the network through the bulk. In considering the c-axis conductivity, we must consider the transmission both through the bulk of one plane to its neighbors as well as from the network of chiral quasi-one dimensional channels of one to the bulk of its neighbors. For the former, according to the above, no Berry phase induced coherence occurs in the plane. The calculation of the c-axis conductivity $\sigma_c^0$ for the majority of the region is therefore to be calculated by the Kubo formula and is proportional to $(v_{Fc})^2$ and gives the usual magneto-resistance. This is also true for the much smaller normal contribuition from connection of the coherent channels in one layer to the incoherent bulk in the adjacent layers.
 
We may take any direction of the field;  for  $B_x > B_y$,  the relevant chains run along ${\hat{y}}$. The transfer integral has a Peierls phase quantized in relation to $\phi_x \equiv B_x L_y d$. The current is linearly proportional to the $e t_c d$ and so  for an external vector potential $A_c$
\be
\label{jc}
j^{\prime}_c =  \sigma^{\prime}_c A_c = e t_c d ~sin(k^0_c d) \frac{{a}_y}{Lx L_y} \frac{\sin  \pi \frac{\phi_x}{\phi_0}}{\pi \frac{\phi_x}{\phi_0}} A_c.
\ee
$ \phi_x = B_x d L_y,~ a_y \approx \xi_{\bot} L_x$ is the area occupied by the channels carrying currents in the plane in the $y$-direction.  The current per-chain is then $\propto \xi_{\bot}$. The resistivity is then approximately $(\sigma^0_c +  \sigma^{\prime}_c)^{-1}$ giving an oscillatory magneto-resistivity to the background resistivity of about
$\approx \sigma^{\prime}_c/(\sigma^0_c)$.
Using for this ratio, just the area occupied by the chains to the total area gives $\xi_{\bot}/4 L_x$, of $O(10^{-2})$. The experimental results are similar to this value, (P.Moll-private communication, Dec. 2025).
 
 Each of these observations summarized in (ii) are shown to follow qualitatively but quantitatively only within an order of magnitude, from the model and calculations above. 
 
{\it (ix) Summary and concluding remarks}: This paper starts with re-capitulating the work of others that looking for coherence through the bulk of the planes beyond the diffusion length is doomed. The idea of Berry phase coherence along chains is found plausible by an estimate of loss-length in chiral chains. Such chains are shown to be promoted due to the energy of coherence between different layers overcoming the energy to creating such chains. Periodic phase shift of the propagation of fermions between the planes by the field then leads to the oscillations of the c-axis magneto-resistivity. Its loss at perpendicular fields that produce order of a flux quantum over the area of the planes is explained. The truly astonishing feature that only field $B_x$ or $B_y$ is relevant for oscillations and the switch at $B_x/B_y$ also follows. 

The numerical estimates made in this paper have  only order of magnitude validity.  The calculations assume that the samples have a short mean-free path compared to sample size but large compared to lattice constants, which is true.
 Only the simplest Fermi-surface has been used, without respect to the geometry of small Fermi-surfaces on a hexagonal Brillouin zone that is expected. This may be justified because the arguments and calculations
are based on general principles and over length scales much larger than lattice constants. This is consistent with the fact that the experiments are a manifestation of macroscopic phase coherence. But the actual details will affect the magnitude of the quantized flux magneto-resistivity by affecting Eq. (\ref{ep}).

 It is useful to ask what can be predicted in future experiments. 
Some are obvious and follow from the experiments already done and physical effects due to coherence learnt from Josephson, that a finite voltage across the sample through an insulating barrier should produce an ac-current with frequency proportional to the voltage and Shapiro-like steps, and that there should be coherent effects in the other two direction between two bulk samples with a weak barrier between them for a field in the plane of each.  There should be hysteresis at the cross-over near $B_x/B_y =1.$
 The crucial test of the developments above would be to directly observe the coherent chiral channels predicted by looking for alternating magnetic fields arranged in strips on the bulk surface using scanning squid or other instruments. 
It would be also be nice to have experiments and calculations to judge their possibility on underdoped cuprates and other twisted bi-layer graphene and other materials in which loop-currents are concluded on the basis of other experiments.

 {\it Acknowledgements}: I thank Philip Moll for discussions and correspondence and to him and Joerg Schmalian for comments on a draft of the manuscript.\\
\bibliography{BerryPhasev2.bbl}

\end{document}